# Non-tidal Coupling of the Orbital and Rotational Motions of Extended Bodies

James H. Shirley

*Jet Propulsion Laboratory, California Institute of Technology, Pasadena, CA, USA*

The orbital motions and spin-axis rotations of extended bodies are traditionally considered to be coupled only by tidal mechanisms. The orbit-spin coupling hypothesis supplies an additional mechanism. A reversing torque on rotating extended bodies is identified. The torque effects an exchange of angular momentum between the reservoirs of the orbital and rotational motions. The axis of the torque is constrained to lie within the equatorial plane of the subject body. Hypothesis testing to date has focused on the response to the putative torque of the Martian atmosphere. Atmospheric global circulation model simulations reveal that an episodic strengthening and weakening of meridional overturning circulations should be observable and is diagnostic in connection with the triggering of Martian planet-encircling dust storms. Spacecraft observations obtained during the earliest days of the 2018 Martian global dust storm document a strong intensification of atmospheric meridional motions as predicted under this hypothesis. We review implications for atmospheric physics, for investigations of planetary orbital evolution with rotational energy dissipation, and for theories of gravitation.

## I. INTRODUCTION

The circulations of planetary atmospheres exhibit marked spatiotemporal variability. Giant storms intermittently appear and disappear within the atmospheres of the outer planets [1-4]. Planet-encircling dust storms occur on Mars in some years, but not in others [5]. On Earth, coherent but poorly understood atmospheric oscillations give rise to multi-year cycles of wet and dry seasons over widespread areas [6-7]. In general, the underlying causes of seasonal, inter-annual, decadal, and longer-term atmospheric variability are poorly understood. Despite considerable investments of resources and time, the predictability of weather and climate on Earth remains limited [8]. Our best numerical models for terrestrial weather prediction currently show very little forecast skill beyond about 2 weeks in the future [9].

Our current understanding of weather and climate variability on Mars compares favorably with that for the Earth. Sub-seasonal time scale forecasts for episodes of large-scale atmospheric instability on Mars are available now for the years 2020-2030 [10]. The Martian global dust storm of 2018, which occurred five Mars years after the previous such storm, was forecasted several years in advance [5, 11-13]. Mars atmospheric global circulation model simulations including orbit-spin coupling [12-13] reproduce the historic record of global dust storm occurrence and non-occurrence on Mars since 1920 with a success rate approaching 80% [14].

These advances are due to incorporating new physics. *Orbit-spin coupling* [15] has twice been introduced within Martian atmospheric global circulation models for hypothesis testing [12-

13]. Table I provides a timeline and list of milestones accomplished in this effort. We here focus on two of the most critical metrics for gauging the legitimacy of new physical hypotheses. We consider:

1: The improvements, if any, afforded in the level of agreement between calculations and observations, and

2: The extent to which new physical hypotheses describe and predict behaviors and processes that have previously not been observed or have been considered to be unpredictable.

We first briefly touch on quantitative aspects (Section II). The coupling equation, its predictive statements, and its effects are then presented and described. In Section IV we describe key results from atmospheric global circulation modeling that yield a diagnostic observable. This section is followed by an overview of recent spacecraft observations that confirm the occurrence within the Mars atmosphere of the predicted intensification of the large-scale circulation during the earliest days of the Martian global dust storm of 2018 [16].

The level of agreement presently achieved between calculations and observations (i.e., between statistical and numerical modeling outcomes and the historic record of Martian dust storm occurrence) is detailed in Sections V and VI. The orbit-spin coupling hypothesis has passed all the tests to which it has thus far been subjected. Section VII describes a method for identifying future episodes of large-scale atmospheric instability on Mars. The technique achieves sub-seasonal resolution in time. We detail a sample forecast (and its outcome) from June and July of 2020. Section VIII provides additional context and discusses implications, while conclusions are detailed in Section IX.

**Table I**. Timeline and milestones achieved in prior investigations. Publications, in the left column, are numbered from P1-P8, while key results (at right) are numbered from R1-R16.

| Prior Work | Principal Findings |
|---|---|
| P1: Shirley, J. H., Solar System Dynamics and Global-scale dust storms on Mars, *Icarus* 251, 128, 2015 | R1. **Discovery** of correlations linking historic Martian global dust storms (GDS) with variations in Mars orbital angular momentum with respect to inertial frames |
| | R2. First **published forecast** calling for a GDS in 2018 |
| P2: Shirley, J. H., Orbit-spin Coupling and the Circulation of the Martian Atmosphere, *Planetary & Space Science* 141, 1-16, 2017 | R3. **Derivation** of the **coupling equation** and demonstration of **quantitative sufficiency** |
| | R4. **Prediction**: Orbital variations drive cycles of intensification and relaxation of atmospheric circulations |
| P3: Shirley, J. H., and M. A. Mischna, Orbit-spin Coupling and the Interannual Variability of global-scale dust storm occurrence on Mars. *Planetary & Space Science* 139, 37-50, 2017 | R5. First **formal statistical test** of the circulatory intensification-relaxation prediction of the orbit-spin coupling hypothesis |
| | R6. Second published forecast calling for a GDS in 2018 |
| P4: Mischna, M. A., & J. H. Shirley, Numerical Modeling of Orbit-spin Coupling Accelerations in a Mars General Circulation Model: Implications for Global Dust Storm Activity, *Planetary & Space Science* 141, 45-72, 2017 | R7. **Hypothesis testing** employing numerical simulations of an atmospheric circulation with orbit-spin coupling. **Confirmation** of the **prediction** of driven cycles of circulatory intensification within the modified GCM, claiming **proof of concept** |
| | R8. **Improved agreement with observations**: First-ever year-by-year **replication of observed planetary-scale atmospheric anomalies**, without the need to pre-condition state variables within the model |
| | R9. Third published forecast calling for a GDS in 2018 |
| | R10. Identification of a **diagnostic observable**: Intermittent cycles of intensification and relaxation of **meridional overturning circulations** |
| P5: Newman, C. E., C. Lee, M. A. Mischna, M. I. Richardson, and J. H. Shirley, An initial assessment of the impact of postulated orbit-spin coupling on Mars dust storm variability in fully interacive dust simulation. *Icarus* 31, 649-668, 2019 | R11. Second GCM investigation demonstrating **proof of concept**. The inclusion of orbit-spin coupling accelerations dramatically inproves the model's skill at predicting GDS and non-GDS years compared to a model without forcing |
| | R12. Fourth published forecast calling for a GDS in 2018 |
| P6: Shirley, J. H., C. E. Newman, M. A. Mischna, & M. I. Richardson. Replication of the Historic Record of Martian Global Dust Storm Occurrence in an Atmospheric General Circulation Model, *Icarus* 317, 197-208, 2019 | R13. **Improved agreement with observation**s: The MarsWRF GCM, with orbit-spin coupling, **reproduces the historic record of Martian GDS** with a success rate of **77%**. |
| P7: Shirley, J. H., A. Kleinböhl, D. M. Kass, L. J. Steele, N. G. Heavens, S. Suzuki, S. Piqueux, J. T. Schofield, and D. J. McCleese, Rapid Expansion and Evolution of a Regional Dust Storm in the Acidalia Corridor During the Initial Growth Phase of the Martian Global Dust Storm of 2018, *Geophysical Research Letters* 46, e2019GL084317, 2019 | R14. **Real-time observation of predicted effects**: The regional -scale "triggering storm" that initiated the 2018 global dust storm was powered-up by an intensified meridional overturning circulation. Spacecraft observations unambiguously record and resolve the **diagnostic observable** for orbit-spin coupling |
| P8: Shirley, J. H., R. J. McKim, J. M. Battalio, & D. M. Kass, Orbit-spin Coupling and the Triggering of the Martian Planet-encircling Dust Storm of 2018, *Journal of Geophysical Research-Planet*s 125, e2019JE006077, 2020 | R15. All historic Martian global dust storms are shown to be associated with dynamically and statistically defined **torque episodes**. |
| | R16. Sub-seasonal time resolution is achieved for hindcasting and for routine **forecasting of intervals of atmospheric instability** on Mars for the years 2020-2030 |

## II. PRELIMINARIES: ANGULAR MOMENTUM IN THE SOLAR SYSTEM

While solar system total angular momentum is conserved, the orbital angular momenta of the Sun and planets individually (with respect to the solar system barycenter, by convention the origin of the solar system inertial frame) exhibit considerable variability with time [5, 17]. Orbital angular momentum is exchanged between the various members of the solar system family on an ongoing and continuous basis. This exchange has until recently been considered to be without significance in connection with the excitation of geophysical variability.

The orbit-spin coupling hypothesis (Section III) identifies and quantifies a weak coupling of the orbital and rotational motions of extended bodies. The order-of-magnitude comparisons of Table II shed light on quantitative aspects of relevance for Mars. Here we see that the *orbital* angular momentum of Mars is >7 orders of magnitude larger than the *rotational* angular momentum of the planet. The rotational angular momentum is in turn about 8 orders of magnitude larger than the angular momentum of the Mars atmosphere [18]. These comparisons draw attention to the fact that even a very tiny exchange of momentum between the orbital and rotational reservoirs may potentially be of considerable geophysical significance. Under the orbit-spin coupling hypothesis, the atmosphere of Mars participates in just such an exchange.

**Table II.** Angular momenta: Representative examples from the solar system.

| | |
|---|---|
| Solar System Total | $3.15 \times 10^{43}$ kg m$^2$ s$^{-1}$ |
| Orbital revolution of Jupiter | $1.90 \times 10^{43}$ kg m$^2$ s$^{-1}$ |
| Rotation of the Sun | $1.92 \times 10^{41}$ kg m$^2$ s$^{-1}$ |
| Solar Barycentric Revolution | < 0 to >$4.60 \times 10^{40}$ kg m$^2$ s$^{-1}$ |
| Orbital revolution of Earth + Moon | $2.68 \times 10^{40}$ kg m$^2$ s$^{-1}$ |
| Orbital revolution of Mars | $3.53 \times 10^{39}$ kg m$^2$ s$^{-1}$ |
| Rotation of Mars | $1.91 \times 10^{32}$ kg m$^2$ s$^{-1}$ |
| Atmospheric circulation of Mars | ~ $10^{24}$ kg m$^2$ s$^{-1}$ |

Quantitative sufficiency is discussed in more detail in [P2] and [P4], where constraints on possible levels of momentum exchange from observations of planetary motions are also described.

## III. THE COUPLING MECHANISM

Figure 1 illustrates barycentric (orbital) revolution and spin-axis rotation in a notional two-body system (such as the Earth and Moon, or the Sun and Jupiter, when considered in isolation).

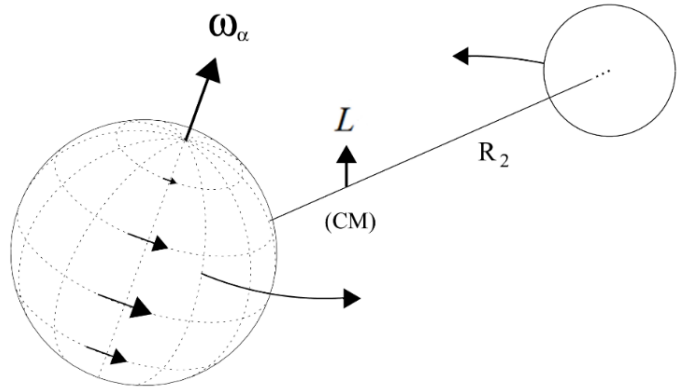

**FIG. 1**. System diagram for axial rotation and orbital revolution in a 2-body system. The curved arrows represent the orbital trajectories of a subject body (at left), and its companion, as they revolve about the center of mass (CM) or barycenter of the pair. ***L*** is a vector representation of the angular momentum of the orbital motion; its direction is normal to the orbit plane. **R** denotes an orbital radius extending from the body center to the system barycenter (here, labeled only for the companion body, i.e., **R**$_2$). The axial rotation (or spin) of the subject body is represented by the angular velocity vector **ω**$_a$.

The orbit-spin "coupling term acceleration," derived in [P2], takes the following form:

$$\boldsymbol{a}_c = -c\,(\dot{\boldsymbol{L}} \times \boldsymbol{\omega}_a) \times \mathbf{r} \tag{1}$$

Here $\dot{\boldsymbol{L}}$ (or $d\boldsymbol{L}/dt$) represents the time rate of change of the orbital angular momentum of the subject body with respect to the solar system center of mass (or barycenter), while the axial rotation of the subject body (with respect to the same inertial coordinate system) is represented by the angular velocity vector **ω**$_a$, as in Fig. 1. **r**, not illustrated in Fig. 1, denotes a position vector, with origin at the body center, as resolved for some specific instant of time, in a rotating, Cartesian, body-fixed coordinate system. The leading multiplier $c$ is a unitless scalar coupling efficiency coefficient, which is constrained by observations to be quite small [P2]. Further discussion of the nature and role of $c$ is deferred to the end of this Section.

The expression on the right side of equation 1 has temporal units of s$^{-3}$. As in [P2], to obtain units of acceleration, we simply integrate with respect to time over an interval of 1 s. Calculated vector component magnitudes are unchanged; numerical values output from equation (1) may thereafter be employed directly for numerical simulations.

In an isolated 2-body system, as illustrated in Fig. 1, orbital angular momentum remains constant. In such a case, the time derivative $\dot{\boldsymbol{L}}$ must necessarily vanish. In $n$-body systems, however, as discussed and illustrated in [P1], and noted earlier in Section II and Table II, subject body orbital angular momenta with respect to inertial coordinates (i.e., with reference to the $n$-body system barycenter) typically vary with time. The Sun, for instance, may gain and lose the equivalent of the total orbital angular momentum of the Earth-Moon system (Table II), over time intervals of a decade or two [17].

Figure 2 illustrates both the input to, and the output from, equation (1), over an interval of 15 yr, for the case of Mars. The rate of change of the orbital angular momentum ($d\boldsymbol{L}/dt$) (Fig. 2a) represents the principal source of variability for the coupling term, since the spin angular velocity $\omega_a$ does not exhibit appreciable variability over short periods of time. Figure 2a also serves to illustrate the incommensurability of cycle times of $dL/dt$ and the Martian year. The variable phasing gives rise to inter-annual variability in the sign and magnitude of the coupling with respect to the seasonal cycle of the Mars year.

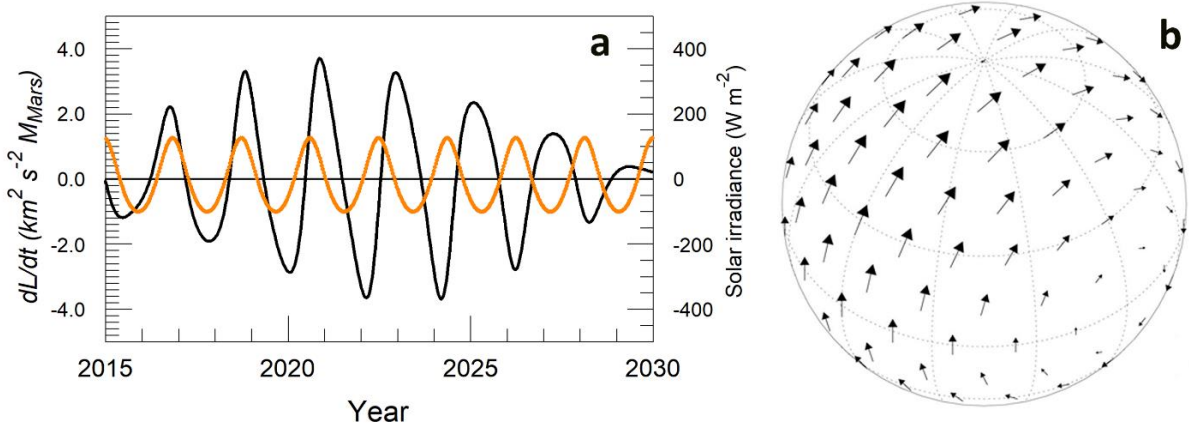

**FIG. 2**. (a) *dL/dt* for Mars in the years 2015-2030, together with the solar irradiance received at Mars (orange curve). The latter is included to illustrate the phasing of the *dL/dt* waveform with respect to the annual cycle of the Mars year. (b) Vector representation of the accelerations imparted by (1) over the surface of an extended body. The lengths of the displayed vectors are proportional to their magnitude. Latitude and longitude grid lines at 30° intervals are shown for reference.

A global view of the accelerations specified by equation (1) is provided in Fig. 2b. Here the north pole of the subject body is near the top of the figure, with latitude and longitude grid lines shown at 30° intervals. The subject body rotates through (or "beneath") the pattern shown, such that the direction of the acceleration, at any given location, cycles in azimuth over ~1 day. The global pattern of accelerations is in some ways similar to the force diagram for a classical belt and pully system; we recognize that the global pattern of accelerations constitutes a torque about an axis lying in the equatorial plane. Due to the frequent reversal in sign of *dL/dt* (Fig. 2a), no large secular precession results from this torque (at least for Mars). Importantly, at the zero crossings, when the *dL/dt* waveform approaches and transitions through zero values, the accelerations of Fig. 2b diminish and disappear, to re-emerge subsequently with reversed directions. Numerical simulations [P4] confirm that momentum is cumulatively added to atmospheric motions, spinning up the atmospheric circulation, during times leading up to and following the extrema of the *dL/dt* waveform. The relaxation (or spin-down) phase occurs in proximity to the zero crossings as illustrated in Fig. 2a.

The torque of equation (1) bears *no functional similarity to tides*. Tides (to be discussed more fully below in Section VIII) have an inverse 3$^{rd}$ power dependence on distance to the disturbing body, while no such dependence is seen in (1). Tides raised on extended bodies may be resolved into (locally) vertical and horizontal components, whereas the acceleration of (1) is everywhere tangential to spherical surfaces. At Mars, calculated peak accelerations due to (1) were found to be larger than the largest (solar) gravitational tidal acceleration by more than 3 orders of magnitude [P2].

The leading coefficient $c$ of equation (1) is in some ways similar to the coefficient of friction $\mu$ first introduced by da Vinci [19], as it operates on and represents a fractional portion of a dynamical quantity [P2]. It is likewise similar in acting as a placeholder for a potentially large catalog of as-yet poorly understood physical interactions likely taking place on molecular or smaller scales. $c$ may help characterize the proportion of the orbital momentum engaged in the excitation of geophysical variability. At Mars, a value of $c=5.0 \times 10^{-13}$ was found to give rise to an appreciable atmospheric response [P4].

A central goal of past (and future) investigations has been (and will be) to determine whether improved agreement between atmospheric numerical modeling outcomes and observations may be attained through the use of a nonzero value of $c$ in equation (1). We return to this topic in Section VIII below.

The torque of equation (1) has not previously been recognized or envisioned, and does not currently appear, either in prior treatments of Newtonian dynamics, or in applications of modern relativistic theories of gravitation.

## IV. GLOBAL CIRCULATION MODEL SIMULATIONS WITH ORBIT-SPIN COUPLING YIELD A DIAGNOSTIC OBSERVABLE

The Mars atmosphere is a complex, nonlinear system, circulating above a planet with quite large topographic variability. A substantial percentage (~30%) of the mass of the atmosphere cycles into and out of seasonal ice caps on both poles. The large eccentricity (0.09) of the planetary orbit gives rise to strong variability of solar heating over the Mars year. Dust storms of various sizes occur every year on Mars. These are most often observed during the southern summer "dust storm season," which is centered roughly on the time of perihelion. In some years, but not in others, regional-scale dust storms grow and coalesce to become planet-encircling in scale [P1, P7, P8].

It is difficult to confidently visualize in advance the consequences of the addition of the orbit-spin coupling accelerations to this dynamic system. The accelerations vary with time and from place to place. Numerical experiments with global circulation models have accordingly been employed to search for patterns of atmospheric behaviors attributable to the accelerations that may be relevant to the problem of Martian global dust storm occurrence.

Two experiments utilizing the MarsWRF Global Circulation Model [20-21] were undertaken for hypothesis testing [P4, P5]. In each case, comparisons were made between unforced model "control runs" and dynamically forced model runs of identical duration. Orbit-spin coupling accelerations were included within the dynamical core of the GCM in forced-model runs. In [P4], to best isolate the effects of the coupling on atmospheric motions, atmospheric dust heating (an important effect at Mars) was omitted. In [P5], radiatively-active dust was added back within the simulations. [P4] considered forced changes in the large scale circulation and in near-surface wind speeds, while simulated atmospheric temperature changes were investigated in [P5]. Despite the exploratory nature of these initial experiments, in both studies, atmospheric conditions favorable for GDS occurrence were reproduced in most of the years in which such storms actually occurred [P4, P6]. We display in Fig. 3 a diagnostic observable, identified in these investigations, which may be characterized as a driven intensification (followed by relaxation) of meridional overturning circulations.

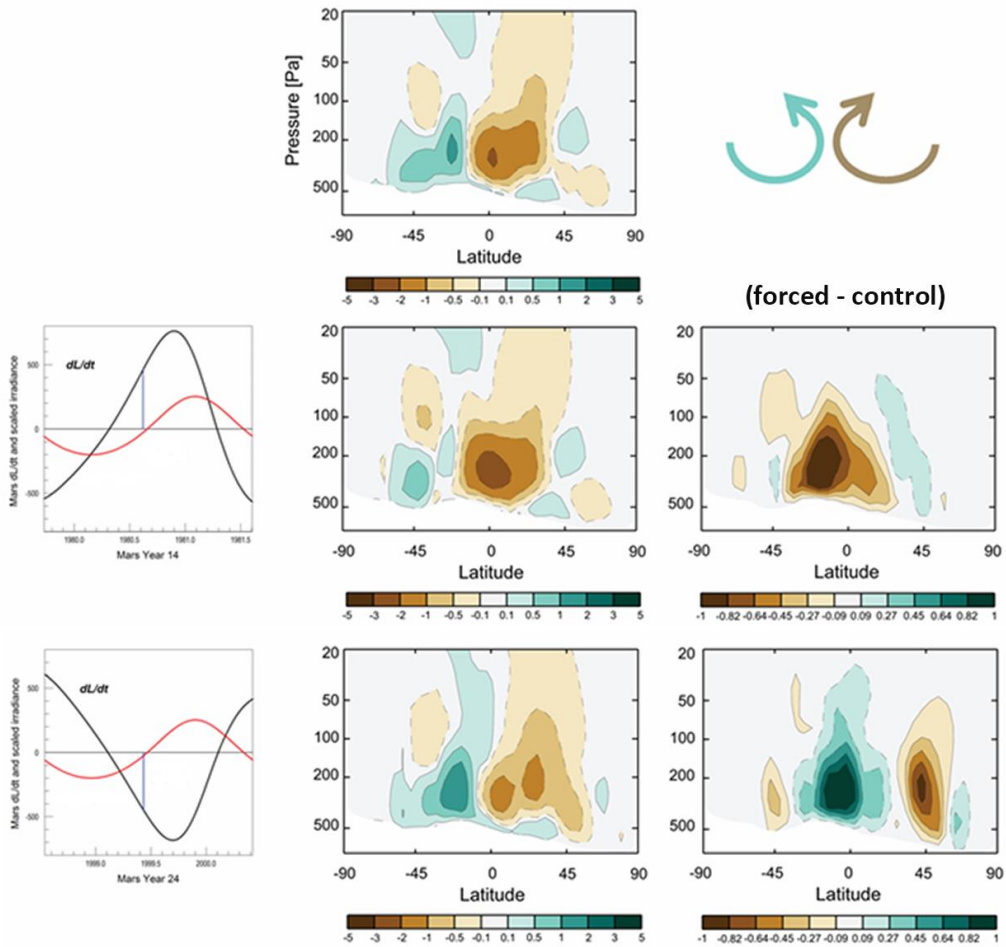

**FIG. 3**. MarsWRF global circulation model zonal mean cross-sections of simulated atmospheric meridional motions as perturbed by the addition of orbit-spin coupling accelerations under two opposed-polarity forcing conditions. Units of flow are $10^9$ kg s$^{-1}$. The vertical axis for the streamfunction plots is atmospheric pressure in Pascals (Pa). The 20 Pa level is typically found at an altitude of ~30 km in the Mars atmosphere. Zonal averaging over a period of ~30 Martian days was employed to generate these plots. Top: Unforced (control run) streamfunction plot showing paired clockwise (brown) and counterclockwise (green) circulation cells. Middle row: When Mars is gaining orbital angular momentum (see forcing function *dL/dt*, at left), the perturbed atmospheric flows (middle), and streamfunction difference plot (forced-control) show a clockwise overturning circulation strengthened by ~20%. Bottom row: When Mars is losing orbital angular momentum (see the forcing function *dL/dt*, at left), the perturbed atmospheric flows (middle), and streamfunction difference plot (forced-control) show a strengthened *counterclockwise* circulation of similar magnitude.

Cross-sections of the Mars atmosphere, extending in latitude from pole to pole, with meridional wind speeds averaged over all longitudes, are presented in the center column of Fig.

3. The top-center panel illustrates a control run simulation showing paired northern (clockwise) and southern (counterclockwise) meridional circulation cells. This twin-cell circulation pattern is observed near the times of the equinoxes on Mars. In both hemispheres, in middle latitudes, air near the surface flows towards the equator, where it is subsequently lofted high in the atmosphere. The circulation is completed by air descending to the surface in higher latitudes. A similar circulation is found above equatorial and middle latitudes on Earth, where it is known as the Hadley Circulation.

Investigators have long suspected that an intensification of the Martian meridional overturning circulation (MOC) could contribute to the growth of global dust storms [22]. The top center panel ("control run") of Fig. 3 illustrates contributing factors. First, an increase in near-surface equatorward wind speeds will plausibly lift more dust from the surface. Stronger upwelling above equatorial regions may thereafter loft the entrained dust to higher altitudes, where dust-induced warming of the atmosphere may further strengthen the circulation in a positive feedback loop. However, the underlying physical processes that could drive MOC intensification in the earliest days of such storms have until recently been obscure.

The second and third rows of Fig. 3 illustrate the changes in the simulated Martian MOC resulting from the addition of orbit-spin coupling accelerations to the MarsWRF GCM [P4]. The middle row of plots illustrate the MOC at times when Mars is gaining orbital angular momentum (left panel) at the season depicted. The center panel represents the forced circulation (which may be compared with the control run immediately above). The panel at right shows the differences in the two, in the sense (forced-control). Here we see that a strengthening (by ~20%) of the clockwise circulation (in brown colors) has occurred (note the difference in scales for the middle and right columns).

The third row of plots of Fig. 3 illustrates consequences for the meridional overturning circulation when Mars is *losing* orbital angular momentum at the season depicted. Strengthening of the *counterclockwise* circulation cell is simulated under these conditions, as indicated by the extensive area plotted in green in the difference plot (at lower right).

Seven of the nine historic Martian global dust storms investigated in [P4] occurred under conditions similar to those displayed in the middle row of panels of Fig. 3. Conversely, the forcing conditions shown in the bottom row of plots of Fig. 3 were found to be unfavorable for the occurrence of perihelion-season GDS. In such cases, the intensified counterclockwise flow tendency interferes destructively with the normal seasonal development of MOC flows during the dust storm season [P2, P4]. (Several other factors linked with the phasing of the *dL/dt* waveform with respect to the annual cycle are also found to be inimical to GDS occurrence. For instance: Years when the *dL/dt* waveform transitions through zero values near the midpoint of the dust storm season are likewise unfavorable for GDS occurrence [P3, P4, P8]).

The above considerations lead us to identify the episodic intensification (and relaxation) of meridional overturning circulations as a *diagnostic observable*.

# V. ANOMALOUS INTENSIFICATION OF THE MARTIAN MERIDIONAL OVERTURNING CIRCULATION OBSERVED AT THE START OF THE GLOBAL DUST STORM OF 2018

The Mars Climate Sounder (MCS) instrument [23] on board the Mars Reconnaissance Orbiter (MRO) spacecraft [24] has obtained limb-sounding atmospheric radiance measurements at Mars for more than 6 Mars years. MCS is a passive 9-channel radiometer viewing the atmosphere with 21 detectors covering altitudes from 0-80 km. The radiances measured are inverted to yield atmospheric temperatures and dust and water ice aerosol profiles [25-26]. Figure 4 shows latitudinal cross-sections of nighttime MCS temperatures (panels a-b) and dayside dust loading (extinction per km; panels c-d). Panels a and c represent seasonal-normal conditions observed before the inception of the GDS of 2018, while panels b and d provide the corresponding data fields for times about 10 Martian days (or "Sols") later, during the peak phase of the 2018 "triggering storm" [P7].

The Martian meridional overturning circulation is indicated by superimposed curved arrows in Fig. 4a. The illustrated trajectories represent the equatorward near-surface winds, the ascending branches of the cells above the equator, and the subsidence in higher latitudes that completes the circulation. We employ temperatures from the night side of the planet to reveal the adiabatic compressional heating in the middle atmosphere (~40 km to ~60 km altitudes) to greatest effect [P7]. (Direct solar heating of the atmosphere on the dayside produces a similar but more complex temperature field).

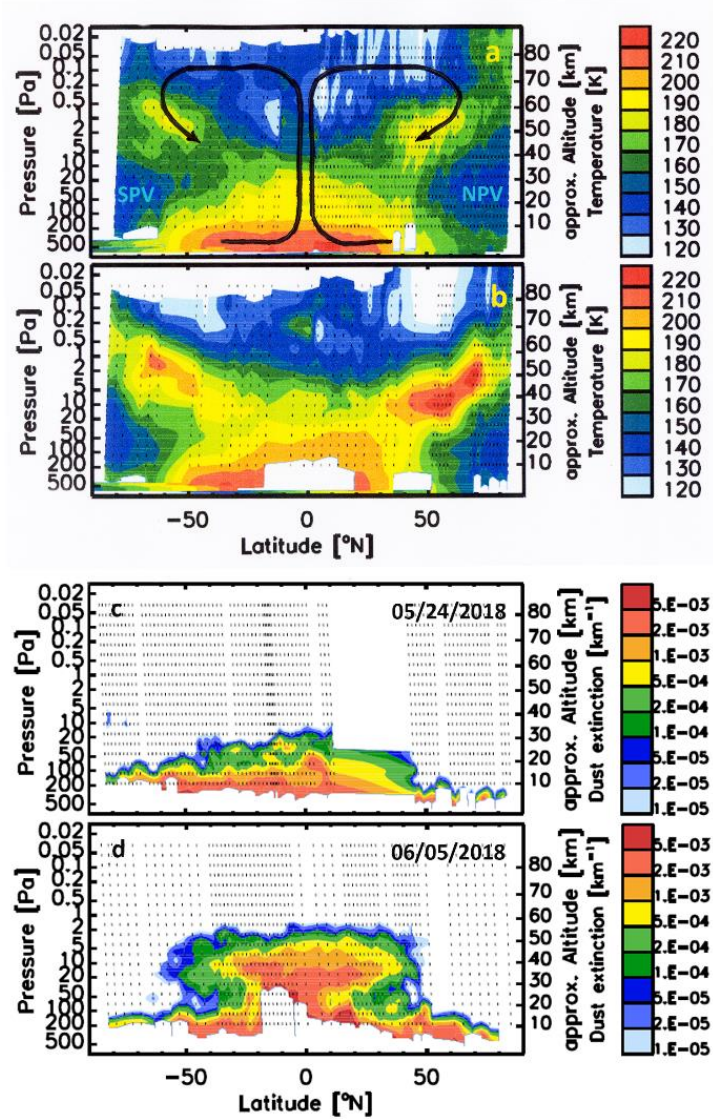

**FIG. 4**. Mars Climate Sounder (MCS) atmospheric cross-sections of temperature (a-b) and dust extinction (c-d) as a function of latitude for times before (a, c) and during (b, d) the triggering regional dust storm of early June 2018 [P7]. Subspacecraft longitudes at times of equator crossing ranged between 4° W and 51°W. Hash marks indicate the latitudes and altitudes of MCS retrievals of atmospheric properties. a) Seasonal-normal nighttime temperatures prior to triggering storm inception. SPV=South Polar vortex, NPV=North Polar vortex. The overturning circulation is indicated by the superimposed curved arrows. b) Nighttime temperatures ~10 Sols later, showing adiabatic warming of ~30 K in the descending branches of the MOC. c) Atmospheric dust loading prior to the triggering storm. d) Atmospheric dust loading ~10 Sols later. Strong upwelling in the intensified meridional overturning circulation has entrained dust to peak altitudes approaching 60 km.

In Fig. 4b we see that temperatures in the middle atmosphere, near the points of the arrowhead symbols of Fig. 4a, increased by ≥30 K in the ~10 Sol interval separating the two plots. The observed increase in adiabatic heating is an unambiguous signature of a strengthened meridional overturning circulation.

Figures 4c and 4d illustrate atmospheric dust loading prior to and during the triggering storm of June 2018 [P7]. The pre-storm dust distribution (Fig. 4c) shows dust layer peak altitudes ≤30 km at all latitudes. In Fig. 4d we note that dust entrained in the intensified MOC has been lofted to altitudes approaching 60 km over an area extending ~100° in latitude.

A detailed discussion of the evolution of the triggering storm shown in Fig. 4 is provided in [P7], where many additional details may be found. We note in passing here 1) that rapid and widespread dust lofting to high altitudes (as in Fig. 4d) has not been seen by orbiting spacecraft in prior Mars years lacking GDS, and 2) that alternate hypotheses for dust lofting to high altitudes were evaluated (in [P7]) but could not explain the scale and rapid development of the storm as observed. With reference to the simulations plotted in Fig. 3, we note that warmer temperatures at altitude (Fig. 4b) in the northern hemisphere indicate greater initial strengthening of the clockwise cell, just as indicated in the center row of plots of Fig. 3 [also see P7].

The intensification of the meridional overturning circulation observed at the start of the GDS of 2018 closely duplicates the key features and morphology of the diagnostic observable previously identified in atmospheric simulations including orbit-spin coupling (Fig. 3). The short timescale of this sequence (≤10 Sols) [also see P8] places the triggering storm events [P7] squarely in the category of "weather" rather than "climate." Choosing the simplest and most direct explanation presently available, we conclude: Orbit-spin coupling torques evidently perturb the weather on Mars.

## VI. IMPROVED AGREEMENT WITH OBSERVATIONS: THE HISTORIC RECORD OF GLOBAL DUST STORM OCCURRENCE ON MARS

Unmodified Martian global circulation models (without adjustable tuning parameters, and lacking the orbit-spin coupling accelerations) show little intrinsic inter-annual variability [27-28], and exhibit no forecast or hindcast skill. Global circulation models incorporating the torque reproduce the historic record of years with and without GDS in 17 of 22 cases [P5], a rate that differs from stochastic forcing model outcomes at the 99% level [P6]. It is clear that the agreement of numerical modeling outcomes with observations is substantially improved by the addition to our models of orbit-spin coupling accelerations [P4, P5, P6].

## VII. TORQUE EPISODES AND SUB-SEASONAL FORECASTING

All of the Martian global dust storms of the historic record have occurred under one or the other of the following two forcing conditions:
1) GDS tend to occur at times when orbit-spin coupling torques are peaking near the middle of the dust storm season (Fig. 5a), and
2) GDS also tend to occur near the times when orbit-spin coupling torques are changing most rapidly (Fig. 5b).

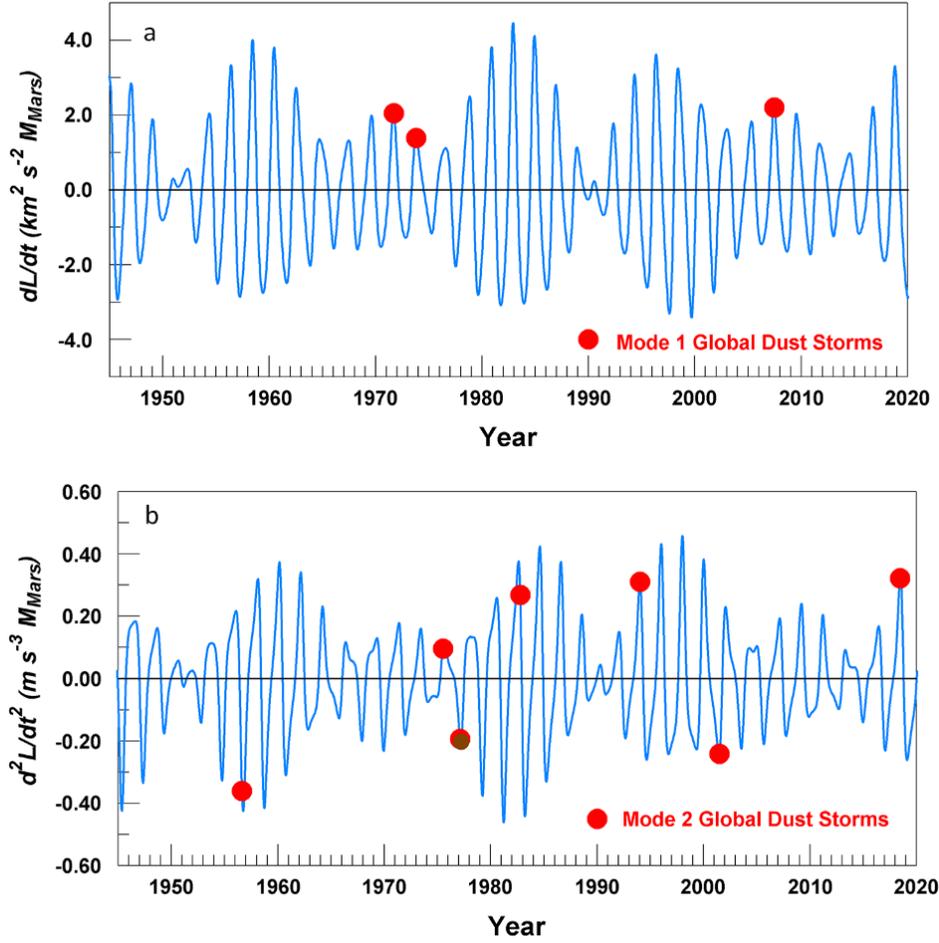

**FIG. 5.** Occurrence times of known Martian global dust storms (1945-2020) with respect to $dL/dt$ (a) and to its time derivative, $d^2L/dt^2$ (b) [P8]. The Mode 1 category includes the GDS of 1971, 1973, and 2007. The Mode 2 category includes the recent 2018 event, along with the historic events of 1956, 1975, 1977, 1982, 1994, and 2001.

      An extended statistical analysis of relationships of the dynamical waveforms of Fig. 5 and the historic record of Martian GDS occurrence since 1870 is found in [P8]. The analysis yields criteria for identifying past and future intervals of increased large-scale atmospheric instability and thus times when dust storm activity is favored. The time intervals thus identified are termed "torque episodes." Results of this analysis for the dust storm season of 2020 are illustrated in Fig. 6. While positive extrema of both the $dL/dt$ and $d^2L/dt^2$ waveforms are in evidence, during the second half of the Mars year, only the $d^2L/dt^2$ peak meets the phasing criteria [P8] for identification as a torque episode. As indicated by the area highlighted in yellow in Fig. 6, the method singles out the period between $L_s=201°$ (15 May 2020) and $L_s=247°$ (28 July 2020) as an interval favorable for Martian dust storm occurrence. ($L_s$, the aerocentric longitude of the Sun, is a standard measure of the progression of the annual cycle of seasons on Mars. $L_s=0°$ corresponds to the time of the Martian vernal equinox).

A large regional-scale dust storm occurred on Mars in late June and early July 2020 [29]. The timing of the observed storm with respect to the time interval of the previously identified torque episode is indicated by the arrowhead symbol in Fig. 6. The phasing of this recent storm is in-family with the phasing of historic Mode 2 GDS as illustrated in Fig. 5b.

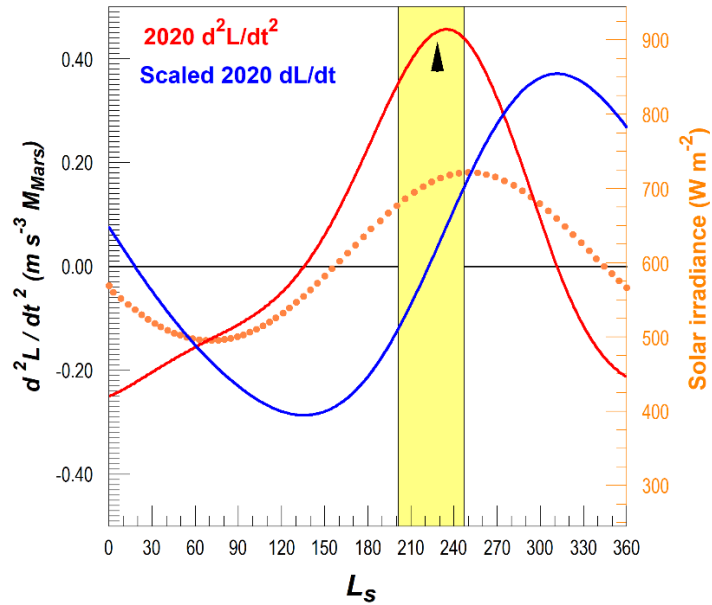

**Figure 6**. The Martian torque episode of 2020 (shaded in yellow) [after Fig. 11 of P8]. Dynamical waveforms (signed $z$ components of $dL/dt$ in blue, and $d^2L/dt^2$ in red) are illustrated for the 2020 Martian dust storm season. The amplitude of the $dL/dt$ waveform has been scaled by a factor of $10^{-8}$ for plotting. Dotted symbols represent the annual cycle of solar irradiance. The arrowhead symbol indicates the initiation of the large regional dust storm of June-July 2020 [29].

The example of Fig. 6 illustrates the approach and the potential for sub-seasonal timescale forecasting of intervals of large-scale atmospheric instability afforded under the orbit-spin coupling hypothesis. The Martian dust storm season may be identified as the interval (as shown in Fig. 6) wherein the solar irradiance is $\geq \sim 600$ Wm$^{-2}$ [5]. The torque episode (identified by the yellow bar in Fig. 6) spans an interval of 72 Sols (Martian days), while the dust storm season by comparison spans an interval of 294 Sols. No corresponding sub-seasonal timescale forecasting capability is presently available for anticipating changes in future terrestrial atmospheric motions and corresponding weather and climate [8-9].

No firm conclusions regarding the coincidence in time of the regional dust storm and the torque episode displayed in Fig. 6 may yet be drawn. Regional-scale dust storms are frequently observed at this season on Mars [30]. As this is written, the relevance and utility of torque episodes can only be considered to be well established in connection with historic global-scale dust storms [P8].

Martian torque episodes for the years 2020-2030 are tabulated and illustrated in [P8]. Only one future torque episode during this interval is considered likely to give rise to a future global-scale dust storm. The identified highest likelihood forecast interval extends from 24 October 2025 to 28 January 2026 [P8].

## VIII. DISCUSSION

Agreement between theoretical predictions and observations is the single most crucial metric for gauging the viability and legitimacy of new physical hypotheses. The explanation of the excess perihelion precession of Mercury [31] by calculations employing general relativity [32] is a textbook example. As summarized in Section VI, the agreement between numerical modeling outcomes and observations is unquestionably improved by including orbit-spin coupling accelerations within global circulation models. While this achievement differs qualitatively from the classic example of the explanation of the excess perihelion precession of Mercury, it is arguably similar in kind.

The ability to describe and predict physical behaviors and processes that have previously not been observed (or have been considered to be unpredictable) is a second metric for gauging the viability and success of physical hypotheses. Einstein's 1915 prediction of the bending of starlight in the Sun's gravitational field [33], and its subsequent confirmation via observations by Eddington and others [34], is an archetypal example. The observation within the Mars atmosphere of the predicted intensification of the meridional overturning circulation (Section V), although less spectacular, is qualitatively similar to the historic example of Einstein's prediction, followed by confirming observations, of the gravitational bending of starlight. In both cases, detailed observations of an infrequent natural event confirmed theoretical predictions made some years in advance of the event.

Additional testing of the hypothesis, employing natural laboratories other than the Mars atmosphere, is now called for. Below we make note of some additional implications and consequences, and highlight certain opportunities for further investigation.

### A. Implications for Studies of Planetary Atmospheres

The governing partial differential equations describing the evolution of atmospheric weather and climate exhibit a sensitive dependence on initial conditions [8-9, 35-37], which is a hallmark of chaotic dynamics. The circulation of the Earth's atmosphere is thereby considered to be fundamentally chaotic [8-9]. The atmospheres of the Earth and Mars are described by the same primitive equations, and thus chaotic dynamics is an expected consequence for the circulation of the atmosphere of Mars as well. Analyses of Martian atmospheric observations reveal diagnostic indications of chaotic dynamics [38], particularly during the Martian dusty season.

Sensitive dependence on initial conditions leads inevitably to the conclusion that details of weather cannot accurately be predicted, even in principal, much beyond 2 weeks in the future [36]. This far-reaching conclusion however rests on the underlying assumption that *all relevant physical processes have been included in the governing equations of the physical model*. We contend instead that the physical system postulated under the chaotic dynamics paradigm is *incomplete*. By adding (and withdrawing) momentum to (and from) atmospheric motions, orbit-spin coupling introduces an additional, dynamical source of atmospheric variability that lies outside the scope of the conventional paradigm for describing atmospheric variability.

Thus, in order to employ a physically complete system description, in future atmospheric numerical modeling, we must necessarily add equation (1) to our list of governing equations. This is, in essence, what was done in [P4 and P5] to achieve the unprecedented multi-decadal hindcast success rates reported in [P6].

Including the coupling term accelerations within numerical global circulation models for the Earth may lead to an improved understanding of the origins of atmospheric "natural variability" (also termed "internal variability"), which is not at present well represented in terrestrial GCMs. We note in passing that recent investigations of terrestrial atmospheric predictability have uncovered "a potentially serious problem with climate models" [39-40], termed the "signal-to-noise paradox" [39], which appears to call into question the paradigm of a sensitive dependence on initial conditions for the Earth's atmosphere. The predictable component [41] of the real atmosphere is found to greatly exceed that for individual atmospheric model runs, which exhibit the expected chaotic effects, over widespread regions and on all time scales [39-41].

Orbit-spin coupling torques experienced by the Earth are larger by a factor of ~5 than those calculated for Mars [Appendix A]. Thus, there is a clear and pressing need for terrestrial investigations of similar nature and scope to those listed in Table I.

## B. Some Implications for Geophysics

Physics plays a unique role (among natural science disciplines) in defining the fundamental underlying "rules of the game" for other disciplines. Precepts based on physical observations and theories may become enshrined as criteria for defining what is and is not acceptable, and viable, in formulating new hypotheses in other disciplines. One such precept has been challenged [P2], and is called into question, under the orbit-spin coupling hypothesis. This is the postulate of the *independence of orbital and rotational motions*.

Wollard [42] expresses this dictum in the following way:

*"Kinematically, the motion of the Earth as a whole can be represented as the resultant of a translation and a rotation in an indefinite number of ways. The particular one that is most advantageous from a dynamical point of view is a representation as the resultant of a translation of the Earth as a whole with the velocity of the center of mass, and a rotation about an axis through the center of mass. These two component motions are **dynamically independent of each other**…"* (emphasis added)

This precept may have its origins in Newton's point mass approximation [43], which has for centuries enabled us to calculate and model planetary motions without considering the rotation states of the gravitating bodies. Corollary to this precept is the widely held assumption among geophysicists that the rotation states of extended bodies and their constituent particles may be rigorously modeled as *closed systems* within which conservation of momentum applies [44]. (Tidal torques, discussed below, comprise a notable and much-studied exception to this rule).

If we assume that the rotation of the Earth (and its constituent parts) may be treated as a closed dynamical system, then conservation of momentum dictates that an increase in the angular momentum of one component, such as the atmosphere, must be accompanied by a decrease of the angular momentum of some other component, such as the Earth's mantle or the liquid core.

Solutions to a number of classical problems have proven to be quite elusive under this paradigm. The decade fluctuations of the Earth rotation [45-47], for instance, are so large as to imply that a coupling of the core and mantle must be involved in their excitation [48-49]. However, no consensus has emerged regarding the physical interaction that actually accomplishes the required core-mantle coupling [50-51]. The excitation mechanism for the geodynamo, which also likely involves core-mantle coupling, likewise remains mysterious [52-53], despite many decades of observation and focused modeling investigations.

Further progress may be possible, with these presently intractable problems, providing we reject the closed-system assumption, and allow for the action and effects of the external torque acting on the Earth as specified by equation (1). The independence of orbital and rotational motions, and the linked closed-system assumption for studies of planetary rotation, in our view, should now be recognized as paired simplifying assumptions that have outlived their utility for geophysical investigations requiring high precision.

### C. Tides and Tidal Torques: *Spin-orbit coupling* vs. *Orbit-spin coupling*

The tide-raising forces are widely considered to represent the only gravitational mechanism operating within the solar system dynamical environment that is capable of imparting relative motions to the constituent particles of extended bodies. In Newtonian theory, the tides represent the gradient of the inverse-square force of attraction spanning the diameter of a subject body. They accordingly exhibit an inverse $3^{rd}$ power dependence on the distance (**d**) separating the centers of the disturbed body and the disturbing body. In relativity, the tides are instead identified with the gradient of the "background geometry of spacetime" [54]. Tidal accelerations obtained using relativistic methods agree with those calculated by Newtonian methods under solar system conditions.

The tide-raising forces, by altering the moments of inertia of extended bodies, effect a coupling of the orbital and rotational motions of those bodies. Numerous tidal effects are accordingly detected in high-resolution time series characterizing the variability of the Earth rotation [45, 50, 55-56]. Tidal torques acting on the equatorial bulges of oblate spinning bodies are responsible for the phenomena of precession and nutation, as for instance in the Earth-Moon system [50, 57]. Precession and nutation are considered to be non-dissipative, and are thus not accompanied by long-term (secular) changes in orbital dimensions or orbital motions.

Inverse-third-power of distance tidal effects are additionally considered to play a pivotal role in a number of astrophysical problems [58-59].

Dissipative tidal friction [50, 60], also known as spin-orbit coupling [61], and tidal heating [62], is thought to exert a dominant influence on the long-term evolution of the Earth-Moon system [63]. In stellar systems, friction associated with tidally induced fluid flow is similarly considered to lead to long-term energy dissipation, with "profound consequences throughout all of astrophysics" [64]. The orbital evolution of gravitationally interacting extended bodies by means of tidal dissipation is thus a topic of great significance in geophysics, in astrophysics, in gravitation, and potentially in cosmology. We are thus motivated to compare the similarities and differences of tidal *spin-orbit coupling* and non-tidal *orbit-spin coupling*.

The two mechanisms are similar, in allowing a transfer of angular momentum between the reservoirs of the orbital and rotational motions. They are likewise similar in positing dissipative losses of energy during this exchange [P2, P8]. In tidal friction, elastic forces within the outer regions of the Earth, together with frictional effects arising from tidal flows in the oceans, are postulated to give rise to a lagged response to the tidal forces. The displaced, non-equilibrium tidal bulge of the Earth exerts a small (orbital) torque on the Moon, leading to a transfer of momentum to the lunar orbital motion. Accompanying losses of rotational energy are brought about by frictional interactions between the ocean and seafloor [45, 50], and within the ocean itself [65]. Under the orbit-spin coupling hypothesis, frictional interactions, for instance between the accelerated flows of the Mars atmosphere and the underlying surface [P8], likewise dissipate energy sourced from the coupled dynamical system. However, in this case, the dynamical system must by definition include the entire solar system, with its vastly larger reservoir of angular momentum (Table II).

Another important difference between the two mechanisms is found in their magnitudes. As already noted in Section III, calculated peak orbit-spin coupling accelerations at Mars are larger (by 3 orders of magnitude) than the tides, which are in turn much larger than the inverse-sixth-power-of-distance dependent accelerations of the tidal torques on Mars' rotation. Similar disparities are expected for other solar system objects with comparable spin rates (cf. Appendix A).

The mechanisms also differ significantly in the character of their variability with time. In the Earth-Moon system, tidal friction produces a slow monotonic transfer of momentum to the lunar orbital motion, and a secular decrease in the rotation rate of the primary [50, 63]. The orbit-spin coupling mechanism, on the other hand, provides *reversing* torques whose phasing (with respect to the annual cycle), and amplitude, vary continuously with time (Fig. 2; Appendix A).

Finally, as documented above and in Table I for Mars, the orbit-spin coupling torques may strongly perturb the atmospheric circulations of extended bodies. No such role has ever been suggested for the torques due to tidal friction.

The classical tidal friction mechanism is not without its difficulties. Investigators have long recognized that an extrapolation of current rates of orbital evolution and dissipation in the Earth-Moon system into the past yields a catastrophically small value for the orbital radius vector of the Moon only ~1.5 Gyr ago [45, 50]. This value is inconsistent with the radiologically-determined age of the Earth of ~4.5 Gyr [67]. This discrepancy is now interpreted as evidence for time varying dissipation, possibly due to changes in the configuration of Earth's ocean basins [68].

Orbit-spin coupling now provides an alternative to tidal spin-orbit coupling for problems of orbital evolution with rotational energy dissipation. We suspect that many effects currently attributed to dissipative tidal processes may instead result from orbit-spin coupling. Additional modeling will be needed to more fully understand the relationships between these mechanisms. An opportunity exists for performing comparative modeling studies that may further illuminate the underlying physics in a wide variety of contexts and situations. In particular, if the use of a nonzero value of $c$ is found to improve the level of agreement between observations and calculations, as in [P4] and [P5], then this may be taken to confirm the relevance of orbit-spin coupling for the problem addressed. This criterion has broad applicability across a range of disciplines. It may be employed for problems ranging from atmospheric time-variability, as in [P4] and [P5], to questions of dynamo excitation [17, 69], and beyond.

### D. Fundamental Physics: More Questions than Answers

One need not possess a deep understanding of the origins of inertia in order to make use of Newton's Laws. The orbit-spin coupling torque of equation (1) may likewise be employed, for practical purposes, in dynamical investigations, in geophysical modeling, and in astrophysical studies, without additional justification. Our scientific curiosity nonetheless demands explanations. Unfortunately, at this early stage, we cannot yet provide a definitive answer to the central question: What is the fundamental underlying process, or mechanism, that allows or enables the weak coupling that is expressed functionally in equation (1)?

We have previously characterized the coupling [P2] as a form of interference, or cross-talk, that weakly couples two dissimilar forms of rotary motion [70] with respect to distant sources. Considerations of Newton's bucket experiment [43] and Mach's Principle [71] have influenced the Author's thinking in this connection. A gedanken experiment highlighting the divergence of (forward) curvilinear motions of rotation and revolution for a single constituent particle of an extended body, as informed by such considerations, was introduced in [P2]. We however recognize that this conceptual framework may not ultimately lead to any deeper understanding.

It may be more fruitful to instead focus our attention (and future theoretical investigations) on the presently unknown locus and/or scale of the coupling. This locus seems likely to reside at molecular or sub-atomic levels (although this is wholly speculative at the time of writing). We conjecture that higher-dimensional gravitational theories may in future more satisfactorily constrain the locus of the coupling, thereby accounting for the findings of Table I and many other currently anomalous observations.

Analogy has been made between the coupling efficiency coefficient $c$ of equation (1) and the coefficient of friction $\mu$ first introduced by Leonardo da Vinci [19]. $c$ currently appears in equation (1) as a scalar placeholder for characterizing, in an approximate but still quantitative fashion, the macroscopic effects of the currently unknown, underlying coupling process. Thus, with reference to the problem of acquiring a deeper understanding, it may be useful to investigate more closely those aspects of the coupling efficiency coefficient $c$ of equation (1) that may be accessible through experimentation.

Numerical values of $c$ obtained and optimized for differing macroscopic components of extended bodies (such as the atmosphere, oceans, and liquid core, in the case of the Earth) are likely to differ, depending upon the physical characteristics of the sub-system examined. Separately resolving the optimal values of $c$ for each sub-system will help constrain the relative magnitudes of macroscopic momentum transfer and exchange, and may thereby illuminate both 1) the interactions between sub-systems, and 2) the details of the locus and scale of the underlying mechanism responsible for the coupling.

Finally, if it is eventually found that nature employs a physics with non-zero $c$ in a wide variety of situations, then $c$ may eventually play yet another role. We conjecture that the determination of $c$ for other natural systems may yield information that can serve to constrain, and discriminate between, competing theories of gravitation. That is, results from new investigations of $c$ may eventually provide information that may allow us to converge on the "most correct" theory or theories of gravitation.

## IX. SUMMARY AND CONCLUSIONS

The orbit-spin coupling hypothesis identifies a reversing torque acting on rotating extended bodies that are members of gravitating $n$-body systems. The axis of the torque is constrained to lie within the equatorial plane of the subject body.

Minute but non-trivial fractional portions of the orbital angular momenta of extended bodies are made available for the excitation of geophysical variability by means of this torque.

The orbit-spin coupling hypothesis has passed all the tests to which it has thus far been subjected. Success has been achieved in meeting two of the most critical metrics for assessing the viability and utility of new physical hypotheses:

1) Improved agreement between calculations and observations: The level of agreement between numerical simulations and historic observations of global dust storms, for the Mars atmosphere, is unquestionably improved by the inclusion of orbit-spin coupling accelerations in global circulation model simulations.

2) Testable predictions: Orbit-spin coupling makes deterministic predictions of cycles of intensification and relaxation of atmospheric circulations. In agreement with prior numerical simulations, a strongly anomalous intensification of the Martian meridional overturning circulation was recorded, by spacecraft observations, at the beginning of the planet-encircling dust storm of 2018.

The orbit-spin coupling hypothesis has in addition led to the first successful years-in-advance forecast of a planetary-scale atmospheric anomaly (the Martian global dust storm of 2018).

Two potentially revolutionary implications have been detailed. The first of these may be stated as follows: Orbit-spin coupling now provides a viable alternative to tidal coupling mechanisms for problems of orbital evolution with rotational energy dissipation.

A second important implication is that the governing equations employed for Martian and terrestrial atmospheric modeling should be amended to include the orbit-spin coupling torque of equation (1). The addition of orbit-spin coupling accelerations to global circulation models for the oceans and atmosphere of the Earth may lead to substantial quantitative improvements in the predictability of terrestrial weather and climate.

**Acknowledgements and Data**

Critical comments on an earlier version of this paper by Jon Giorgini, Peter Read, Jim Murphy, Bruce Bills, and Tim McConnochie are gratefully acknowledged. Algorithms for calculating solar system barycentric orbital angular momentum and its derivatives are described in references [5], [11], and [12]. Basic data for dynamical calculations was obtained from JPL's Horizons system (https://ssd.jpl.nasa.gov/horizons.cgi). Mars Climate Sounder data covering the period of the 2018 global dust event is available on the Planetary Data System. We acknowledge with thanks many discussions with members of JPL's Mars Climate Sounder Science and Operations Teams, including Dan McCleese, David Kass, Armin Kleinböhl, Tim Schofield, Rich Zurek, Shigeru Suzuki, Tina Tillmans, and Jason Matthews. Atmospheric global circulation model simulations were performed at NASA's Ames Research Center, via NASA's High-End Computing Program, with support from JPL's Research and Technology Development Program and NASA's Solar System Workings Program. Portions of this work were performed at the Jet Propulsion Laboratory, California Institute of Technology, under a contract from NASA. Copyright 2020, California Institute of Technology. Government sponsorship acknowledged.

**APPENDIX A: COMPARING THE ORBITAL ANGULAR MOMENTA AND ORBIT-SPIN COUPLING TORQUES ON EARTH AND MARS**

In Section VIII we state that the orbit-spin coupling torques acting on the Earth are significantly larger than those acting on Mars. Below we show a sample calculation and comparison. We will assume that the values of the coupling efficiency coefficient $c$ for Mars and the Earth are identical for this exercise. To illustrate the difference in magnitude, in Fig. A1, we compare the time derivatives of the angular momentum per unit mass ($h$) of the two bodies. The parameter $h$ was chosen to allow comparison of the phasing of the two waveforms on a common scale.

We note immediately in Fig. A1 the larger amplitude of the terrestrial waveform and its approximately monthly modulation due to the presence of Earth's Moon. Due to the orbital motion of the Earth and Moon about their common barycenter, Earth's orbital angular momentum (with respect to the solar system inertial frame) is alternately greater than, and less than, its long-term mean value. The waveform for Mars in Fig. A1 corresponds with that shown earlier (in Fig. 2) of this paper. The peak-to-peak cycle times for both planets are slightly longer than one solar year, as defined for each subject body, due to the ongoing prograde orbital motion of the Sun with respect to the solar system barycenter during one annual revolution of each planet [5].

The true ratio of the amplitudes of the orbit-spin coupling torques on the two bodies is somewhat larger than is indicated in Fig. A1. If we wish to compare accelerations at the surfaces of these planets, due to the factor **r** in equation (1), we must additionally account for the differences in the planetary dimensions. The ratio of the planetary radii is ~1.88 (6.378 km / 3.396 km).

Peak values of the orbit-spin coupling accelerations at the surface of Mars for the period 1920-2030 were calculated in [P4]. An acceleration of $2.2 \times 10^{-4}$ ms$^{-2}$ was obtained when employing a value for the coefficient $c$ of $5 \times 10^{-13}$. A similar calculation, made for the time of the largest positive peak of the terrestrial $dL/dt$ waveform in Fig. A1, returns a value of $1.17 \times 10^{-3}$ m s$^{-2}$, when likewise employing a $c$ value of $5 \times 10^{-13}$. This is larger than the peak orbit-spin coupling acceleration at Mars by a factor of 5. This acceleration may usefully be compared with the peak value of the tidal acceleration of the Moon on the Earth. At the time of closest approach of the Moon to the Earth (perigee), the lunar tidal acceleration at the Earth's surface attains a value of ~ $1.3 \times 10^{-6}$ m s$^{-2}$.

The results of these comparisons once again highlight the need for terrestrial investigations aimed at iteratively constraining the coefficient $c$ for use in modeling the circulation of the Earth's atmosphere (as done for Mars in [P4]).

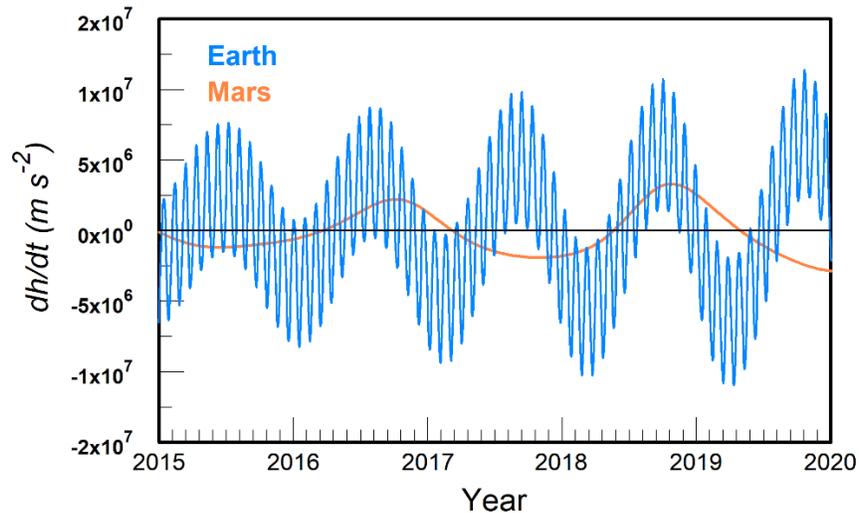

**FIG. A1**. Rates of change of the specific angular momentum $h$ for Earth and Mars (with respect to the solar system barycenter) for the years 2015-2020. The figure provides values of the $z$ component of $dh/dt$ with respect to ecliptic coordinates.